\documentclass[a4paper,11pt]{article}
\usepackage{pos}

\title{Determination of hybrid charmonium meson masses}

\author*{Gaurav Ray}
\author{Craig McNeile}
\affiliation{Centre for Mathematical Sciences, University of Plymouth, Plymouth, UK}

\emailAdd{gaurav.sinharay@plymouth.ac.uk}

\abstract{We report initial results from our study of the masses and decay constants of the lightest multiplet of charmonium-like hybrid mesons. We obtain precise measurements of the $1^{-+}$ state through the use of a variational basis and a large number of configurations at three lattice spacings. We use staggered fermion operators using configurations generated with the HISQ action with 2+1+1 dynamical flavours. The mixing of the vector hybrid with the $J/\psi$ is examined and a preliminary bound on the vector hybrid decay constant is presented.}

\FullConference{%
 The 38th International Symposium on Lattice Field Theory, LATTICE2021
  26th-30th July, 2021
  Zoom/Gather@Massachusetts Institute of Technology
}

\begin{document}
\maketitle

\section{Introduction} \label{se:intro}
Since 2003 a series of resonances have been observed in the charmonium and bottomonium sectors that do not correspond neatly with Quark model states \cite{Belle:2003goh, BaBar:2004oro}. Many competing non-Quark model explanations for these so-called $XYZ$ states have been proposed, including (but not limited to) gluonic hadrons, bound states of more than 3 quarks, and hadronic molecules \cite{Ali:2017jda, Guo:2017jvc, Brambilla2019d}.
A hybrid meson is a meson \emph{with an excited gluonic component} \cite{Mar1999}.
Although hybrid mesons (and the other states we've noted) are exotic in the sense that they fall outside the well-tested Quark model, nothing in principle prevents $\bar{q}qg$ states in QCD.
Hybrid mesons have been studied on the lattice for many decades \cite{campbell_heavy_1988,Lacock:1996ny,Bernard:2003, Ma2019, Bernard1998, Cheung2016a}.
Relevant experimental searches include the upcoming 
PANDA experiment 
at FAIR~\cite{PANDA:2021ozp}, which will search for evidence of gluonic excitations in the hadron spectrum, and GlueX at the Jefferson Lab \cite{Hamdi2019,barucca_panda_2021}.  Accurate lattice QCD calculations of the properties of the hybrid mesons in charmonium may encourage the LHCb collaboration to search for them.

In this paper we present initial results for the mass of the $1^{-+}$ and $1^{--}$ hybrid mesons in charmonium, based on lattice QCD calculations using the staggered formalism. The advantages of using staggered fermions is that the lattice spacing errors are small and the HPQCD collaboration has done many comparison between lattice calculations and the experimental properties of mesons in the charmonium system~\cite{Dowdall:2012ab}. There are some disadvantages of using staggered fermions over other formulations of the Dirac operator. For example, because the staggered correlators have a contribution from the parity partner, the hybrid mesons are the first excited state, 
but in Wilson like formulations the hybrid  meson is the ground state. The initial goal of this project was to check whether it
was possible to get accurate results for the mass of hybrid mesons made from charm quarks using staggered fermions.

Although at this stage in the project we focus on single meson operators, we will eventually need to include two meson operators. There are only a few exploratory calculations of hadronic decays using staggered fermions~\cite{Fu:2012gf,Fu:2016itp}.
For completeness, we briefly discuss the possible hadronic decays of hybrid charm mesons. There are possible string breaking-like decays where the hybrid meson can decay into pairs such as $\overline{D}D$, or two P-wave D mesons, or a P-wave D and a S-wave D, as well as similar channels with charm-strange mesons. There is a selection rule in the heavy quark limit~\cite{Michael:1999ge} (and some quark models~\cite{Page:1996rj}) that a hybrid meson can not decay into two S-wave mesons. It is possible for the hybrid meson to decay to a standard charmonium meson with a light meson and there is a lattice QCD calculation in the heavy quark limit~\cite{McNeile:2002az}.

\section{Interpolating operators for hybrid mesons}

Since hybrids have an excited gluonic component the operators which we would expect to couple to them include the field strength tensor, $F_{ij}^{ab}$, whose components are the chromoelectric and chromomagnetic fields ($\frac{1}{2}\epsilon_{ijk} F_{jk} = B_i$). These are contracted, over the colour indices, with the usual fermion bilinears (so that the quark-antiquark pair is a colour octet).  \\
As we are using staggered fermions we replace the $\gamma$ matrices with phases. These phases, along with the $F_{ij}^{ab}$, determine the quantum numbers of the operators. In the staggered basis the operators also have a taste assignment. Taste breaking effects should drop out in the continuum limit.

We started by considering the lightest hybrid multiplet, as determined by the HadSpec collaboration in \cite{Cheung2016a},
\begin{align}
    &1^{-+} : \epsilon_{ijk}\Bar{\psi}\gamma_{j}B_{k}\psi  &\longrightarrow&  &\gamma_{i}\otimes\gamma_{i} &: \Bar{\chi}\epsilon_{ijk}(-1)^{x_{j}}B_{k}\chi  \\
    &1^{--}_{H} : \Bar{\psi}\gamma_{5}B_{i}\psi  &\longrightarrow&  &\gamma_{5}\otimes\gamma_{5} &: \Bar{\chi}B_{i}\chi \\
    &0^{-+}_{H} : \Bar{\psi}\gamma_{i}B_{i}\psi &\longrightarrow& &\gamma_{i}\otimes\gamma_{i} &: \Bar{\chi}(-1)^{x_{i}}B_{i}\chi  \\
    &2^{-+}_{H} : |\epsilon_{ijk}|\Bar{\psi}\gamma_{j}B_{k}\psi &\longrightarrow& &\gamma_{i}\otimes\gamma_{i} &: \Bar{\chi} |\epsilon_{ijk}|(-1)^{x_{j}}B_{k}\chi \hspace{3mm}.
\end{align}
Note that we have suppressed the colour indices, labelled the states that can mix with conventional charmonium by a $H$ subscript, and given the `spin $\otimes$ taste' assignment for the operators formed from staggered fields $\chi$ on the right.
We have neglected disentangling states due to the reduced cubic symmetry of the lattice, such as  $1^{-+}$ with $4^{-+}$~\cite{Dudek:2009qf}.

The majority of the results from lattice QCD for the masses of hybrid mesons have used clover or Wilson fermions. There has been one calculation by the MILC collaboration~\cite{Bernard:2003} using staggered fermions, which computed the mass of the $1^{-+}$ hybrid meson with light and strange quarks.
The previous MILC calculation used the taste singlet non-local $\rho$ when constructing hybrid operators to minimize taste breaking~\cite{Bernard:2003}. We use the local $\rho$ operator because taste breaking is less of an issue  with charm quarks and local operators can be less noisy than non-local operators.

\section{Simulation Details}

We use the Highly Improved Staggered Quark (HISQ) 
action~\cite{Follana:2006rc} with 2+1+1 sea quarks. The gauge configurations were provided courtesy of the MILC Collaboration~\cite{Bazavov2010a,Bazavov2013a}.
The lattice spacing is set using  $w_0$  with the value calculated by HPQCD~\cite{Dowdall:2013rya}.
In table~\ref{tab:runs} the ensembles used are listed.
We used the MILC code to do the calculations. We modified the existing hybrid operators in the MILC code and tested the correlators against our own 
implementation in the Grid library~\cite{Boyle:2015tjk}.

\begin{table}[ht]
    \centering
    \resizebox{1.0\columnwidth}{!}{%
    \begin{tabular}{|c|c|c|c|c|c|c|c|c|} \hline
      name & size & $a$ (fm) & $m_l/m_s$ & $M_{\pi}L$ & $M_{\pi}$ (MeV) & $am_c^{\mathrm{sea}}$ & $am_c^{\mathrm{val}}$ & configurations \\ \hline
      very coarse & 32$^3$x48 & 0.15088(79) & 1/27 & 3.30 & 131.0(1) & 0.8447 & 0.863 & 1505 \\
      coarse      & 32$^3$x64 & 0.12225(65) & 1/10 & 4.29 & 216.9(2) & 0.628  & 0.650 & 1000 \\
      fine        & 32$^3$x96 & 0.09023(48) & 1/5  & 4.50 & 312.7(6) & 0.440  & 0.450 & 1008 \\
      \hline
    \end{tabular}%
    }
    \caption{The lattice ensembles used. Both the fine and coarse lattices are at heavier than physical pion masses, while the very coarse lattice is at the physical point. The tuned valence charm quark masses are from~\cite{Hatton:2020qhk}}
    \label{tab:runs}
\end{table}

The correlators of hybrid meson operators are typically very noisy. We therefore employ a series of well-established techniques to reduce this noise. We average over multiple time sources and polarisations per configuration. We do variational smearing with local and covariant smearing applied to the quarks.
DeTar and Lee  report on using variational smearing with staggered fermions~\cite{DeTar:2014gla}.
We apply APE smearing on the gauge links in the field strength tensor and in the Gaussian smearing on the quarks.

\section{Results for the mass of the \texorpdfstring{$1^{-+}$}{} hybrid meson}

\begin{figure}[ht]
        \centering
        \captionsetup{width=.7\linewidth}
        \includegraphics[width=0.6\textwidth]{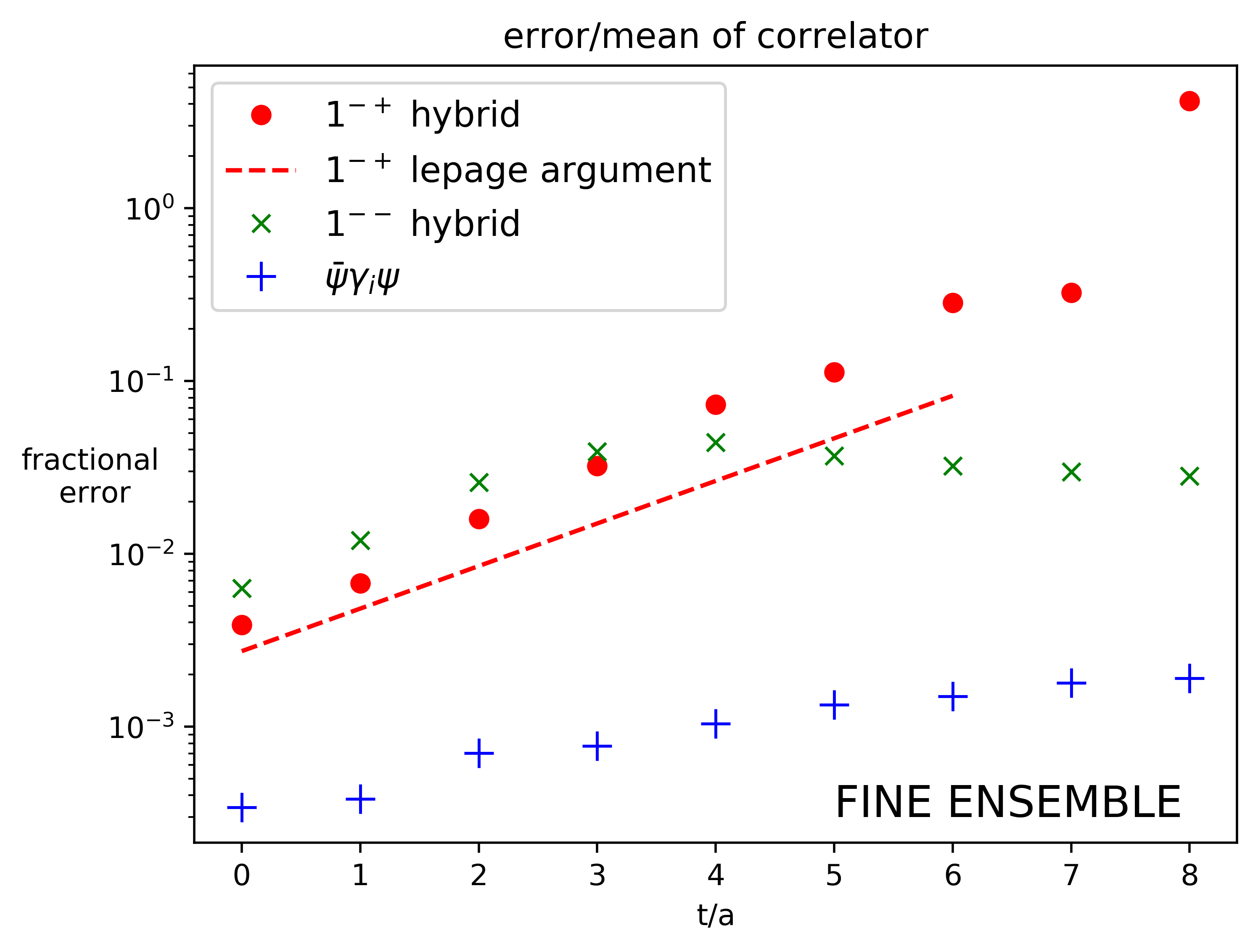}
        \caption{The error/mean of correlators against time for $1^{-+}$, $1^{--}$ hybrid and conventional vector operators on the fine ensemble. We include the predicted fractional error from an argument by Lepage, using the fitted hybrid and pseudoscalar masses. }
        \label{fig:noise}
\end{figure}

In figure~\ref{fig:noise} we plot fractional errors for the $1^{-+}$ hybrid correlator,
$1^{--}$ hybrid correlator, and for comparison the $J/\psi$ correlator with charm quarks on the fine ensemble. The hybrid correlators are considerably more noisy than the conventional vector operator, $\bar{\psi}\gamma_i\psi$, with the signal dying by $t=8$ at the latest.
In figure~\ref{fig:noise} we also compare the fractional error of the $1^{-+}$ hybrid correlator with the estimate from Lepage~\cite{Lepage:1989hd}, which says it should be proportional to $\exp\big(-(M_H-M_{\eta_c})t\big)$.

We use Lepage's python library \texttt{corrfitter}, which implements a suite of functions to fit correlators and matrices of correlators within a Bayesian 
framework~\cite{Lepage2002,lepage_gplepagecorrfitter_2020}.
In the $1^{-+}$ analysis our fitting procedure is as follows. We use a 2 operator basis, with and without smearing, to produce a $2\times2$ matrix of correlators.
We choose a time $t_0$ and use it to generate priors by diagonalising the correlator matrix with the eigenvectors associated to the solution of a Generalized Eigenvalue Problem (GEVP) at $t_0$ and $t_0+1$. The staggered formalism requires a parity partner (PP) state, which oscillates in time, to be included in the fit model. For the $1^{-+}$ channel the quantum numbers of the parity partner are $1^{++}$.
We vary the fit range until the goodness of fit parameters indicate an acceptable fit. Our preliminary results are in table~\ref{tab:fits}.

\begin{table}[ht]
    \centering
    \resizebox{1.0\columnwidth}{!}{%
    \begin{tabular}{|c|c|c|c|c|c|c|c|c|} \hline
      Ensemble & \# tsrc & $t_0$ & range & svdcut & $\chi^2$ per dof & Q & Mass (GeV) & PP Mass (GeV) \\ \hline
      very coarse & 16 & 1 & 1-6 & $3\times 10^{-7}$ & 1.1 & 0.31 & 4.29(11) & 3.89(25)\\
      coarse      & 16 & 1 & 1-5 & $3\times 10^{-5}$ & 1.3 & 0.22 & 4.575(66) & 3.58(28) \\
      fine        & 16 & 2 & 2-8 & $7\times 10^{-4}$ & 0.93 & 0.55 & 4.23(18) & 3.35(48) \\
      \hline
    \end{tabular}%
    }
    \caption{Summary of $1^{-+}$ fits. We define a good fit as having $\chi^2 /$dof $\sim 1$ and a Q value $>0.1$. As discussed in the text `PP' is the Parity Partner state.}
    \label{tab:fits}
\end{table}

\begin{figure}[ht]
    \centering
    \begin{minipage}{0.9\linewidth}
        \centering
        \captionsetup{width=.8\linewidth}
        \includegraphics[width=0.9\linewidth]{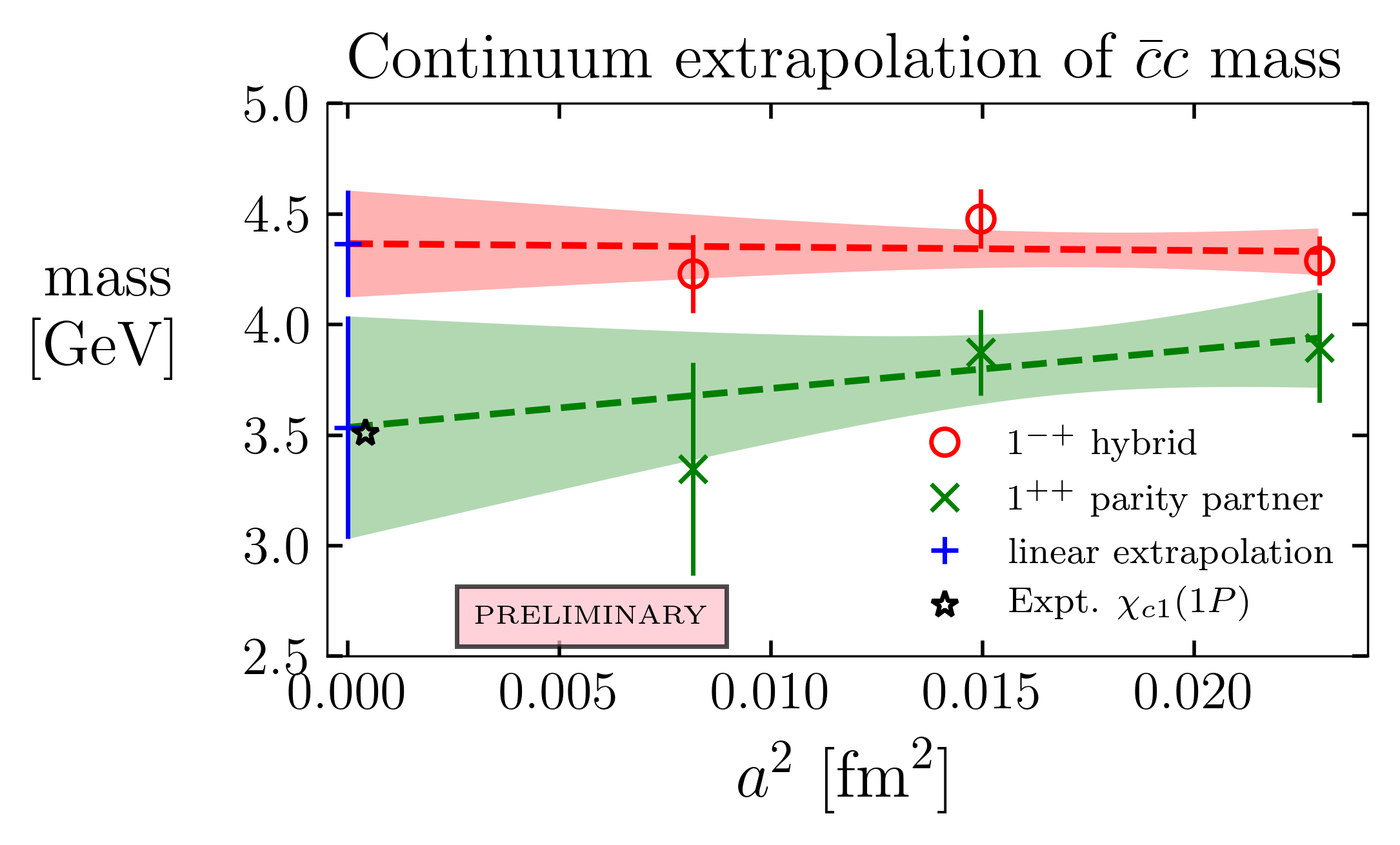}
        \caption{Continuum extrapolation of the mass of the $1^{-+}$ hybrid charmonium meson and the parity partner state. The extrapolated PP mass agrees with experiment albeit with a large uncertainty.}
        \label{fig:onemp_continuum}
        \end{minipage}\hspace{10mm}
\end{figure}

In figure~\ref{fig:onemp_continuum} we show a continuum extrapolation of the mass of the charmonium $1^{-+}$ and parity partner $1^{++}$ state using our results at three lattice spacings. We only use a linear dependence on $a^2$ to do the continnum extrapolation  and neglect any light quark mass dependence or finite volume effects in this preliminary result.
The continuum limit of the mass of the parity partner state agrees with the mass of the $\chi_c(1P)$, which is an important cross-check. In figure~\ref{fig:onemp_summary} we plot our results with the results from the HadSpec collaboration~\cite{Cheung2016a} and Bali et al.~\cite{Bali2011}. We also include some of the physical decay 
thresholds in figure~\ref{fig:onemp_summary}, as briefly discussed in section~\ref{se:intro}.

\section{Results for the properties of the \texorpdfstring{$1^{--}$}{} hybrid meson}


\begin{figure}
     \centering
     \begin{minipage}{0.57\linewidth}
        \centering
        \vspace{5mm}
        \captionsetup{width=.9\linewidth}
          \includegraphics[width=1.0\linewidth]{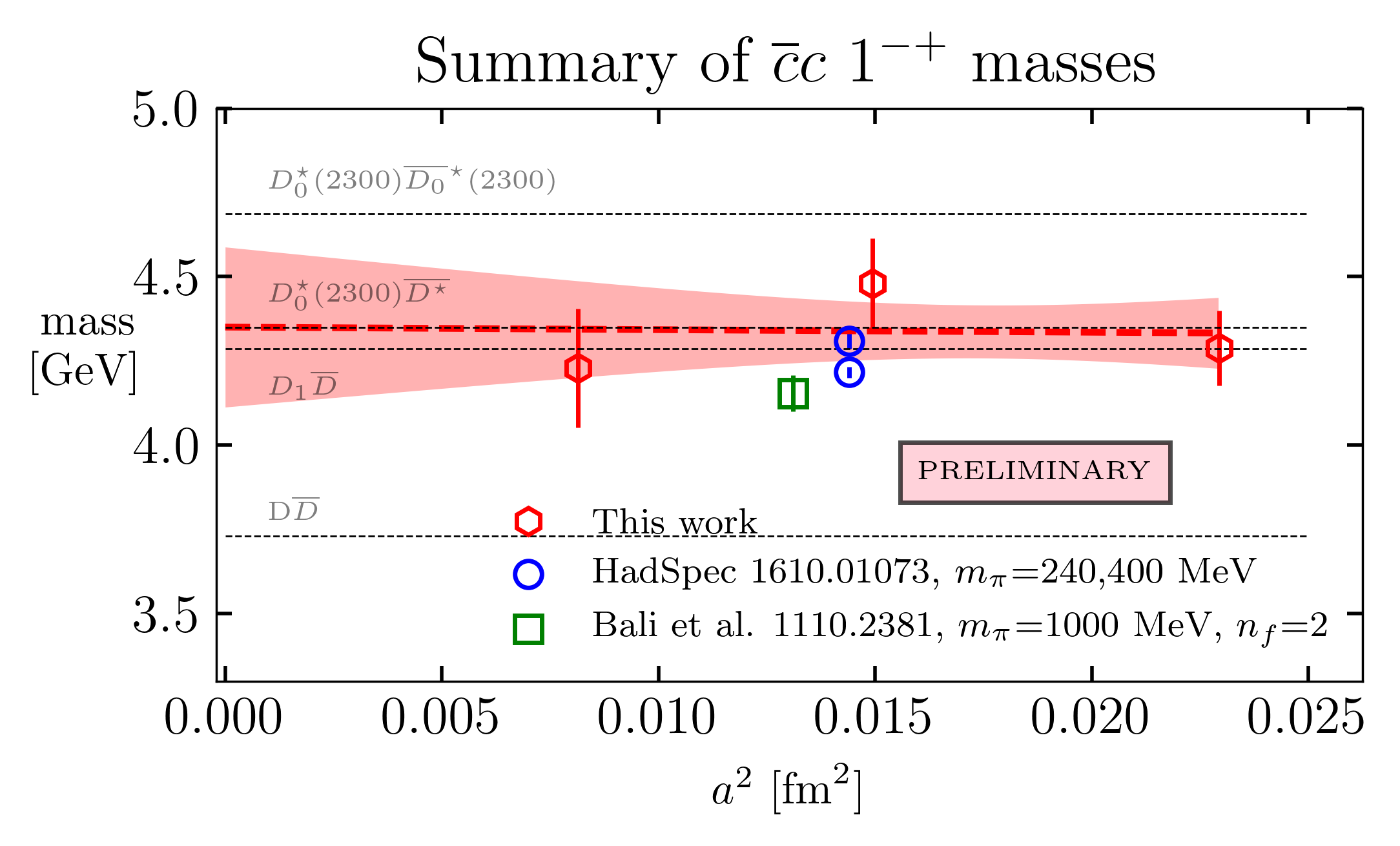}
          \caption{Summary plot for the $1^{-+}$ mass as a function of the square of the lattice spacing, including some of the previous determinations of the mass by other groups \cite{Cheung2016a,Bali2011}} 
         \label{fig:onemp_summary}
     \end{minipage}\hspace{3mm}
     \begin{minipage}{0.4\linewidth}
            \centering
             \includegraphics[scale=0.6]{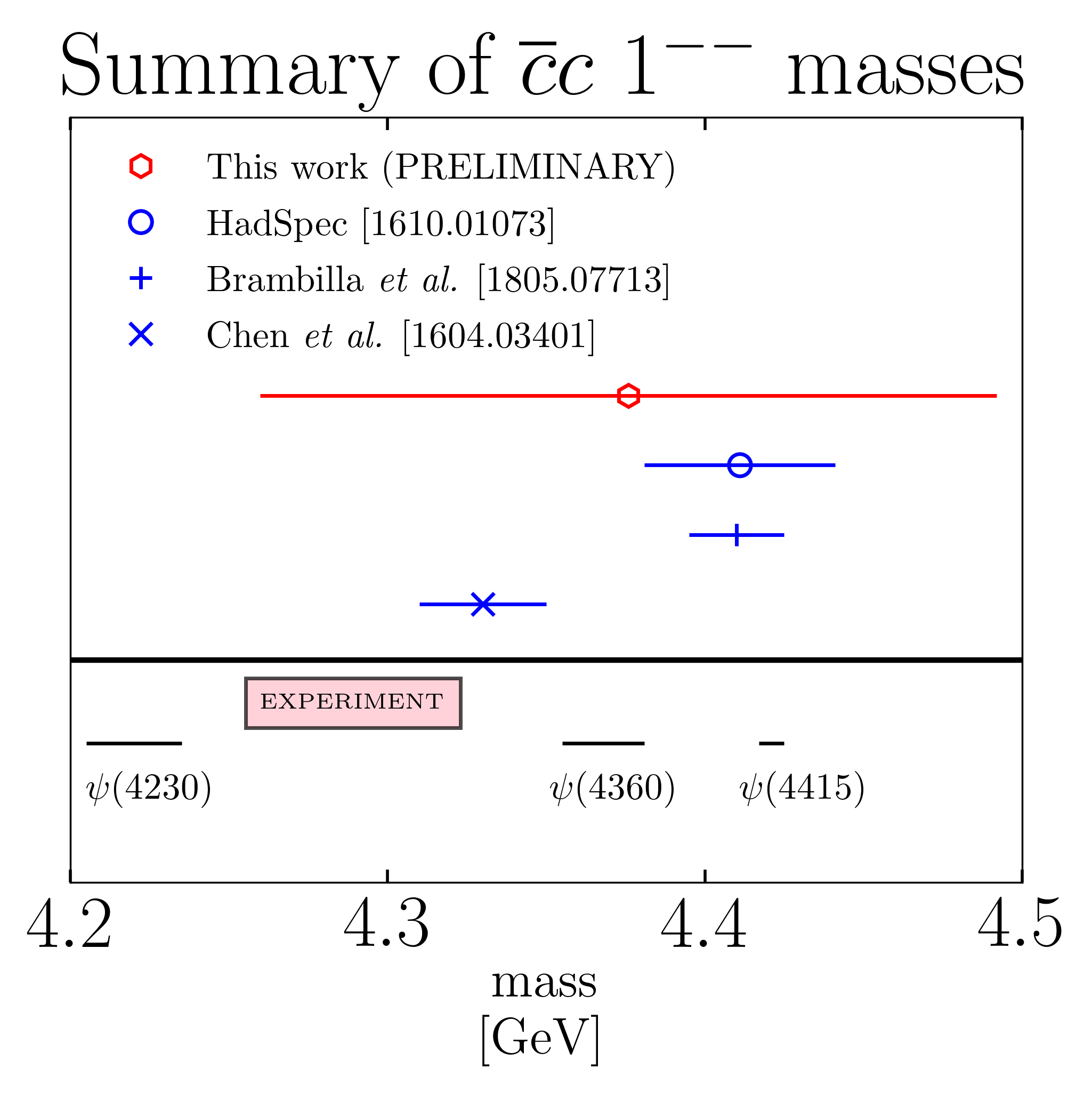}
             \caption{Comparison of our determination of the $1^{--}$ hybrid mass at 0.09 fm from a 4-by-4 fit to the results of three other groups, and three resonances the PDG lists as `estabished'\cite{Cheung2016a,Brambilla2019, Chen2016}. }
            \label{fig:onemm_summary}  
     \end{minipage}
\end{figure}

There are many unexplained resonances with $J^{PC} = 1^{--}$, where charmonium hybrid mesons may occur. See the review by 
Brambilla et al.~\cite{Brambilla2019d}. For example it has been speculated that the $\psi$(4230) is a hybrid meson, but confirmation will require accurate lattice QCD calculations. 

To examine the $1^{--}$ hybrid state we perform a GEVP analysis with a 4 operator basis: standard vector operator, the hybrid operator, and their smeared and local counterparts.
Here we assume the hybrid state is the \emph{second} excited state, after the $J/\psi$ and $\psi(2S)$. We therefore expect the extraction of its properties to be more difficult than in the $1^{-+}$ case. Our initial results are shown in table~\ref{tab:jpsi-fine} and a comparison of our $1^{--}$ hybrid mass to other groups is given in figure~\ref{fig:onemm_summary}.

\begin{table}[ht]
    \centering
     \captionsetup{width=.91\linewidth}
    \resizebox{0.95\columnwidth}{!}{%
    \begin{tabular}{|c|c|c|c|c|c|c|c|c|}
         \hline
         state & fit range & $\chi^2$ per dof & Q & mass [GeV] & amplitude & $f$ [MeV] & $\Gamma_{ee}$ [keV] & comment \\
         \hline
         $J/\psi$   &  &  &  & 3.097(17) & 0.16441(26) & 417.5(2.3) & 5.836(36) & concurrent 2x2 fit \\
         $\psi(2S)$ & 6-24 & 0.88 & 0.72   & 3.781(28) & 0.1860(78) & 428(18) & 5.01(42) & with only \\
         $h_c(1P)$  &  &  &   & 3.512(29) & 0.0578(75)  & -- & -- & vector ops \\
         \hline
         Hybrid     & 1-6  & 0.9 & 0.58  & 4.33(16) & 0.086(15)  & -- & --  & 2x2 fit w/only hybrid ops \\
         \hline
         $J/\psi$   & 2-6 & 0.81 & 0.82 & 3.110(18)    &  0.1701(27)  & 431.1(7.2) & 6.20(20) & concurrent 4x4 fit \\
         Hybrid     &  &  &  & 4.38(12) &  0.065(18)   & 9(167)  & 0.002(67) & with hybrid and vector ops\\
         \hline 
    \end{tabular}%
    }
    \caption{Fit results for the conventional charmonia and the hybrid state on the fine ensemble. The $h_c(1P)$ meson is the parity partner state. The masses and decay constants from the 2-by-2 fits agree well with experiment. The 4-by-4 and 2-by-2 masses are consistent, though there is a slight tension between the the $J/\psi$ amplitudes (and therefore the decay constants too). The leptonic decay widths are also shown in the penultimate column. The hybrid leptonic width is small, consistent with zero, and has a large uncertainty, stemming from the sizeable uncertainty in the amplitude.}
    \label{tab:jpsi-fine}
\end{table}

Brambilla et al.~\cite{Brambilla2019d} review the importance of the 
decay constant in probing the properties of $1^{--}$ hybrid mesons.
The decay constants can constrain the leptonic decay width. 
The decay constant, $f_H$, is defined through a matrix element,
\begin{equation}
\langle 0 | \hat{\cal{V}}_i | H \rangle = f_{H} M_{H} \epsilon_{i} \hspace{3mm},
\end{equation}
where $\hat{\cal{V}}$ is the vector current operator. The local vector operator was used so we use the $Z_V$ renormalization factor from~\cite{Hatton:2019gha}. As a check we computed the decay constant of the $J/\psi$ meson (see~\cite{Hatton:2020qhk} for a systematic study of the leptonic decay constants of the $J/\psi$ meson).

From this we can compute the leptonic width of our hybrid vector charmonium state using 
\begin{equation}
    \Gamma (V_{c\Bar{c}} \hookrightarrow e^+ e^- ) = \frac{16\pi}{27} \alpha_{\mathrm{QED}}^2 \frac{f_H^2}{M_H} \hspace{2mm},
\end{equation}
\noindent where $\alpha_{\mathrm{QED}}$ is the electromagnetic coupling at the charm quark mass. Our amplitude for the vector operator into the $1^{--}$ hybrid state is very small. From this amplitude we obtain an upper bound of $70$ eV. A previous calculation of this leptonic width bounded it from above at $40$ eV~\cite{Chen2016}.

\section{Summary and outlook}

We have presented a first continuum extrapolation of the mass of the $1^{-+}$ charmonium hybrid meson from unquenched lattice QCD. Additionally, preliminary results for the mass and leptonic decay constant of the $1^{--}$ hybrid meson at a single lattice spacing have been reported. We are currently computing correlators for $1^{-+}$ operators  with charm quarks on the coarse and fine ensembles with physical pion masses.
Also we are refining our analysis of the 4-by-4 variational analysis of the $1^{--}$ mesons. Results at a finer lattice spacing ($\sim 0.06$ fm) would improve the quality of our continuum extrapolation.
We plan to compute the mass of the $1^{-+}$ hybrid meson with 
bottom quarks~\cite{Ryan:2020iog}  using the extrapolation method with the HISQ action~\cite{McNeile:2011ng}.

\section{Acknowledgements} 

This
work was supported by the UK Science and Technology Facilities Council. The calculations used
the DiRAC Data Analytic system at the University of Cambridge, operated by the University of
Cambridge High Performance Computing Service on behalf of the STFC DiRAC HPC Facility
(www.dirac.ac.uk). This is funded by BIS National e-infrastructure and STFC capital grants and
STFC DiRAC operations grants.

\bibliographystyle{JHEP}
\bibliography{references,craig}

\providecommand{\href}[2]{#2}\begingroup\raggedright\begin{thebibliography}{10}

\bibitem{Belle:2003goh}
{\scshape Belle} collaboration, \emph{{Observation of a new narrow charmonium
  state in exclusive $B^{\pm} \rightarrow K^{\pm}\pi^+ \pi^- J / \psi$
  decays}},  in \emph{{21st International Symposium on Lepton and Photon
  Interactions at High Energies (LP 03)}}, 8, 2003
  [\href{https://arxiv.org/abs/hep-ex/0308029}{{\ttfamily hep-ex/0308029}}].

\bibitem{BaBar:2004oro}
{\scshape BaBar} collaboration, \emph{{Study of the $B \to J/\psi K^- \pi^+
  \pi^-$ decay and measurement of the $B \to X(3872) K^-$ branching fraction}},
  \href{https://doi.org/10.1103/PhysRevD.71.071103}{\emph{Phys. Rev. D}
  {\bfseries 71} (2005) 071103}
  [\href{https://arxiv.org/abs/hep-ex/0406022}{{\ttfamily hep-ex/0406022}}].

\bibitem{Ali:2017jda}
A.~Ali, J.S.~Lange and S.~Stone, \emph{Exotics: {Heavy} pentaquarks and
  tetraquarks},
  \href{https://doi.org/10.1016/j.ppnp.2017.08.003}{\emph{Progress in Particle
  and Nuclear Physics, Vol 63, No 1} {\bfseries 97} (2017) 123}.

\bibitem{Guo:2017jvc}
F.-K.~Guo, C.~Hanhart, U.-G.~Meißner, Q.~Wang, Q.~Zhao and B.-S.~Zou,
  \emph{Hadronic molecules},
  \href{https://doi.org/10.1103/RevModPhys.90.015004}{\emph{Reviews of Modern
  Physics} {\bfseries 90} (2018) 015004}.

\bibitem{Brambilla2019d}
N.~Brambilla, S.~Eidelman, C.~Hanhart, A.~Nefediev, C.P.~Shen, C.E.~Thomas
  et~al., \emph{The {XYZ} states: {Experimental} and theoretical status and
  perspectives},
  \href{https://doi.org/10.1016/j.physrep.2020.05.001}{\emph{Physics Reports}
  {\bfseries 873} (2020) 1}.

\bibitem{Mar1999}
L.A.~Griffiths, C.~Michael and P.E.~Rakow, \emph{Mesons with excited glue},
  \href{https://doi.org/10.1016/0370-2693(83)90680-9}{\emph{Physics Letters B}
  {\bfseries 129} (1983) 351}.

\bibitem{campbell_heavy_1988}
N.A.~Campbell, A.~Huntley and C.~Michael, \emph{Heavy quark potentials and
  hybrid mesons from {SU}(3) lattice gauge theory},
  \href{https://doi.org/10.1016/0550-3213(88)90170-8}{\emph{Nuclear Physics B}
  {\bfseries 306} (1988) 51}.

\bibitem{Lacock:1996ny}
{\scshape UKQCD} collaboration, \emph{{Hybrid mesons from quenched QCD}},
  \href{https://doi.org/10.1016/S0370-2693(97)00384-5}{\emph{Phys. Lett. B}
  {\bfseries 401} (1997) 308}
  [\href{https://arxiv.org/abs/hep-lat/9611011}{{\ttfamily hep-lat/9611011}}].

\bibitem{Bernard:2003}
C.~Bernard, T.~Burch, E.B.~Gregory, D.~Toussaint, C.~DeTar, J.~Osborn et~al.,
  \emph{Lattice calculation of ${1}^{\ensuremath{-}+}$ hybrid mesons with
  improved kogut-susskind fermions},
  \href{https://doi.org/10.1103/PhysRevD.68.074505}{\emph{Phys. Rev. D}
  {\bfseries 68} (2003) 074505}.

\bibitem{Ma2019}
Y.~Ma, W.~Sun, Y.~Chen, M.~Gong and Z.~Liu, \emph{A {Color} {Halo} {Scenario}
  of {Charmonium}-like {Hybrids}},
  \href{https://doi.org/10.1088/1674-1137/ac0ee2}{\emph{arXiv} (2019) 1}.

\bibitem{Bernard1998}
C.~Bernard, T.~Blum, T.A.~DeGrand, C.~DeTar, S.~Gottlieb, U.M.~Heller et~al.,
  \emph{Hybrid mesons in quenched lattice {QCD}},
  \href{https://doi.org/10.1016/S0920-5632(97)00467-2}{\emph{Nuclear Physics B
  - Proceedings Supplements} {\bfseries 60} (1998) 61}.

\bibitem{Cheung2016a}
G.K.~Cheung, C.~O’Hara, G.~Moir, M.~Peardon, S.M.~Ryan, C.E.~Thomas et~al.,
  \emph{Excited and exotic charmonium, {Ds} and {D} meson spectra for two light
  quark masses from lattice {QCD}},
  \href{https://doi.org/10.1007/JHEP12(2016)089}{\emph{Journal of High Energy
  Physics} {\bfseries 2016} (2016) }.

\bibitem{PANDA:2021ozp}
{\scshape PANDA} collaboration, \emph{{PANDA Phase One}},
  \href{https://doi.org/10.1140/epja/s10050-021-00475-y}{\emph{Eur. Phys. J. A}
  {\bfseries 57} (2021) 184}
  [\href{https://arxiv.org/abs/2101.11877}{{\ttfamily 2101.11877}}].

\bibitem{Hamdi2019}
A.~Hamdi, \emph{Search for exotic states in photoproduction at {GlueX}},
  \href{https://doi.org/10.1088/1742-6596/1667/1/012012}{\emph{Journal of
  Physics: Conference Series} {\bfseries 1667} (2020) 2}.

\bibitem{barucca_panda_2021}
G.~Barucca, F.~Davì, G.~Lancioni, P.~Mengucci, L.~Montalto, P.P.~Natali
  et~al., \emph{{PANDA} {Phase} {One}: {PANDA} collaboration},
  \href{https://doi.org/10.1140/epja/s10050-021-00475-y}{\emph{The European
  Physical Journal A} {\bfseries 57} (2021) 184}.

\bibitem{Dowdall:2012ab}
R.J.~Dowdall, C.T.H.~Davies, T.C.~Hammant and R.R.~Horgan, \emph{{Precise
  heavy-light meson masses and hyperfine splittings from lattice QCD including
  charm quarks in the sea}},
  \href{https://doi.org/10.1103/PhysRevD.86.094510}{\emph{Phys. Rev. D}
  {\bfseries 86} (2012) 094510}
  [\href{https://arxiv.org/abs/1207.5149}{{\ttfamily 1207.5149}}].

\bibitem{Fu:2012gf}
Z.~Fu, \emph{{Preliminary lattice study of $\sigma$ meson decay width}},
  \href{https://doi.org/10.1007/JHEP07(2012)142}{\emph{JHEP} {\bfseries 07}
  (2012) 142} [\href{https://arxiv.org/abs/1202.5834}{{\ttfamily 1202.5834}}].

\bibitem{Fu:2016itp}
Z.~Fu and L.~Wang, \emph{{Studying the $\rho$ resonance parameters with
  staggered fermions}},
  \href{https://doi.org/10.1103/PhysRevD.94.034505}{\emph{Phys. Rev. D}
  {\bfseries 94} (2016) 034505}
  [\href{https://arxiv.org/abs/1608.07478}{{\ttfamily 1608.07478}}].

\bibitem{Michael:1999ge}
C.~Michael, \emph{{Quarkonia and hybrids from the lattice}},
  \href{https://doi.org/10.22323/1.003.0001}{\emph{PoS} {\bfseries hf8} (1999)
  001} [\href{https://arxiv.org/abs/hep-ph/9911219}{{\ttfamily
  hep-ph/9911219}}].

\bibitem{Page:1996rj}
P.R.~Page, \emph{{Why hybrid meson coupling to two S wave mesons is
  suppressed}},
  \href{https://doi.org/10.1016/S0370-2693(97)00438-3}{\emph{Phys. Lett. B}
  {\bfseries 402} (1997) 183}
  [\href{https://arxiv.org/abs/hep-ph/9611375}{{\ttfamily hep-ph/9611375}}].

\bibitem{McNeile:2002az}
{\scshape UKQCD} collaboration, \emph{{Hybrid meson decay from the lattice}},
  \href{https://doi.org/10.1103/PhysRevD.65.094505}{\emph{Phys. Rev. D}
  {\bfseries 65} (2002) 094505}
  [\href{https://arxiv.org/abs/hep-lat/0201006}{{\ttfamily hep-lat/0201006}}].

\bibitem{Dudek:2009qf}
J.J.~Dudek, R.G.~Edwards, M.J.~Peardon, D.G.~Richards and C.E.~Thomas,
  \emph{{Highly excited and exotic meson spectrum from dynamical lattice QCD}},
  \href{https://doi.org/10.1103/PhysRevLett.103.262001}{\emph{Phys. Rev. Lett.}
  {\bfseries 103} (2009) 262001}
  [\href{https://arxiv.org/abs/0909.0200}{{\ttfamily 0909.0200}}].

\bibitem{Follana:2006rc}
{\scshape HPQCD, UKQCD} collaboration, \emph{{Highly improved staggered quarks
  on the lattice, with applications to charm physics}},
  \href{https://doi.org/10.1103/PhysRevD.75.054502}{\emph{Phys. Rev. D}
  {\bfseries 75} (2007) 054502}
  [\href{https://arxiv.org/abs/hep-lat/0610092}{{\ttfamily hep-lat/0610092}}].

\bibitem{Bazavov2010a}
A.~Bazavov, C.~Bernard, C.~Detar, W.~Freeman, S.~Gottlieb, U.M.~Heller et~al.,
  \emph{Scaling studies of {QCD} with the dynamical highly improved staggered
  quark action},
  \href{https://doi.org/10.1103/PhysRevD.82.074501}{\emph{Physical Review D -
  Particles, Fields, Gravitation and Cosmology} {\bfseries 82} (2010) 1}.

\bibitem{Bazavov2013a}
A.~Bazavov, C.~Bernard, J.~Komijani, C.~Detar, L.~Levkova, W.~Freeman et~al.,
  \emph{Lattice {QCD} ensembles with four flavors of highly improved staggered
  quarks}, \href{https://doi.org/10.1103/PhysRevD.87.054505}{\emph{Physical
  Review D - Particles, Fields, Gravitation and Cosmology} {\bfseries 87}
  (2013) 1}.

\bibitem{Dowdall:2013rya}
R.J.~Dowdall, C.T.H.~Davies, G.P.~Lepage and C.~McNeile, \emph{{Vus from pi and
  K decay constants in full lattice QCD with physical u, d, s and c quarks}},
  \href{https://doi.org/10.1103/PhysRevD.88.074504}{\emph{Phys. Rev. D}
  {\bfseries 88} (2013) 074504}
  [\href{https://arxiv.org/abs/1303.1670}{{\ttfamily 1303.1670}}].

\bibitem{Boyle:2015tjk}
P.~Boyle, A.~Yamaguchi, G.~Cossu and A.~Portelli, \emph{{Grid: A next
  generation data parallel C++ QCD library}},
  \href{https://arxiv.org/abs/1512.03487}{{\ttfamily 1512.03487}}.

\bibitem{Hatton:2020qhk}
{\scshape HPQCD} collaboration, \emph{{Charmonium properties from lattice
  $QCD$+QED : Hyperfine splitting, $J/\psi$ leptonic width, charm quark mass,
  and $a^c_\mu$}},
  \href{https://doi.org/10.1103/PhysRevD.102.054511}{\emph{Phys. Rev. D}
  {\bfseries 102} (2020) 054511}
  [\href{https://arxiv.org/abs/2005.01845}{{\ttfamily 2005.01845}}].

\bibitem{DeTar:2014gla}
C.~DeTar and S.-H.~Lee, \emph{{Variational method with staggered fermions}},
  \href{https://doi.org/10.1103/PhysRevD.91.034504}{\emph{Phys. Rev. D}
  {\bfseries 91} (2015) 034504}
  [\href{https://arxiv.org/abs/1411.4676}{{\ttfamily 1411.4676}}].

\bibitem{Lepage:1989hd}
G.P.~Lepage, \emph{{The Analysis of Algorithms for Lattice Field Theory}},  in
  \emph{{Theoretical Advanced Study Institute in Elementary Particle Physics}},
  6, 1989.

\bibitem{Lepage2002}
G.P.~Lepage, B.~Clark, C.T.~Davies, K.~Hornbostel, P.B.~Mackenzie,
  C.~Morningstar et~al., \emph{Constrained curve fitting},
  \href{https://doi.org/10.1016/S0920-5632(01)01638-3}{\emph{Nuclear Physics B
  - Proceedings Supplements} {\bfseries 106-107} (2002) 12}.

\bibitem{lepage_gplepagecorrfitter_2020}
P.~Lepage, \emph{gplepage/corrfitter: corrfitter version 8.1.1},  Nov., 2020.
\newblock 10.5281/ZENODO.4281296.

\bibitem{Bali2011}
G.S.~Bali, S.~Collins and C.~Ehmann, \emph{Charmonium spectroscopy and mixing
  with light quark and open charm states from {nF}=2 lattice {QCD}},
  \href{https://doi.org/10.1103/PhysRevD.84.094506}{\emph{Physical Review D -
  Particles, Fields, Gravitation and Cosmology} {\bfseries 84} (2011) }.

\bibitem{Brambilla2019}
N.~Brambilla, W.K.~Lai, J.~Segovia, J.~Tarrús~Castellà and A.~Vairo,
  \emph{Spin structure of heavy-quark hybrids},
  \href{https://doi.org/10.1103/PhysRevD.99.014017}{\emph{Physical Review D}
  {\bfseries 99} (2019) 1}.

\bibitem{Chen2016}
Y.~Chen, W.F.~Chiu, M.~Gong, L.C.~Gui and Z.F.~Liu, \emph{Exotic vector
  charmonium and its leptonic decay width},
  \href{https://doi.org/10.1088/1674-1137/40/8/081002}{\emph{Chinese Physics C}
  {\bfseries 40} (2016) 1}.

\bibitem{Hatton:2019gha}
{\scshape HPQCD} collaboration, \emph{{Renormalizing vector currents in lattice
  QCD using momentum-subtraction schemes}},
  \href{https://doi.org/10.1103/PhysRevD.100.114513}{\emph{Phys. Rev. D}
  {\bfseries 100} (2019) 114513}
  [\href{https://arxiv.org/abs/1909.00756}{{\ttfamily 1909.00756}}].

\bibitem{Ryan:2020iog}
{\scshape Hadron Spectrum} collaboration, \emph{{Excited and exotic bottomonium
  spectroscopy from lattice QCD}},
  \href{https://doi.org/10.1007/JHEP02(2021)214}{\emph{JHEP} {\bfseries 02}
  (2021) 214} [\href{https://arxiv.org/abs/2008.02656}{{\ttfamily
  2008.02656}}].

\bibitem{McNeile:2011ng}
C.~McNeile, C.T.H.~Davies, E.~Follana, K.~Hornbostel and G.P.~Lepage,
  \emph{{High-Precision $f_{B_s}$ and HQET from Relativistic Lattice QCD}},
  \href{https://doi.org/10.1103/PhysRevD.85.031503}{\emph{Phys. Rev. D}
  {\bfseries 85} (2012) 031503}
  [\href{https://arxiv.org/abs/1110.4510}{{\ttfamily 1110.4510}}].

\end{thebibliography}\endgroup

\end{document}